\documentclass[prl,twocolumn,showpacs,superscriptaddress,floatfix]{revtex4}

\usepackage{graphics}  
\usepackage{makeidx}
\usepackage{colordvi}
\RequirePackage{xspace}

\usepackage{relsize}
\def\babar{\mbox{\slshape B\kern-0.1em{\smaller A}\kern-0.1em
    B\kern-0.1em{\smaller A\kern-0.2em R}}}

\def\epem       {\ensuremath{e^+e^-}\xspace}
\def\qqbar {\ensuremath{q\overline q}\xspace}
\def\ccbar {\ensuremath{c\overline c}\xspace}
\def\piz   {\ensuremath{\pi^0}\xspace}
\def\pip   {\ensuremath{\pi^+}\xspace}
\def\pim   {\ensuremath{\pi^-}\xspace}
\def\Kp    {\ensuremath{K^+}\xspace}
\def\Km    {\ensuremath{K^-}\xspace}
\def\KS    {\ensuremath{K^0_{\scriptscriptstyle S}}\xspace} 
\def\Dbar    {\kern 0.2em\overline{\kern -0.2em D}{}\xspace}

\def\Dz      {\ensuremath{D^0}\xspace}
\def\Dzb     {\ensuremath{\Dbar^0}\xspace}
\def\Dp      {\ensuremath{D^+}\xspace}
\def\Dm      {\ensuremath{D^-}\xspace}
\def\Dstar   {\ensuremath{D^*}\xspace}
\def\Dstarp  {\ensuremath{D^{*+}}\xspace}
\def\Dstarm  {\ensuremath{D^{*-}}\xspace}

\def\Bbar    {\kern 0.18em\overline{\kern -0.18em B}{}\xspace}

\def\BB      {\ensuremath{B\Bbar}\xspace} 
\def\Bz      {\ensuremath{B^0}\xspace}
\def\Bzb     {\ensuremath{\Bbar^0}\xspace}
\def\BzBzb   {\ensuremath{\Bz {\kern -0.16em \Bzb}}\xspace}

\def\jpsi     {\ensuremath{{J\mskip -3mu/\mskip -2mu\psi\mskip 2mu}}\xspace}
\def\Y#1S{\ensuremath{\Upsilon{(#1S)}}\xspace}
\def\FourS {\Y4S}
\def\bpsiks     {\ensuremath{\Bz \to \jpsi \KS}\xspace}

\def\Bztodstdst {\ensuremath{\Bz \to \Dstarp \Dstarm}\xspace}
\def\upsbb   {\ensuremath{\FourS \to \BB}\xspace}
\def\mes        {\mbox{$m_{\rm ES}$}\xspace}
\newcommand{\mev}{\ensuremath{\mathrm{\,Me\kern -0.1em V}}\xspace}
\newcommand{\gevc}{\ensuremath{{\mathrm{\,Ge\kern -0.1em V\!/}c}}\xspace}
\newcommand{\mevc}{\ensuremath{{\mathrm{\,Me\kern -0.1em V\!/}c}}\xspace}
\newcommand{\gevcc}{\ensuremath{{\mathrm{\,Ge\kern -0.1em V\!/}c^2}}\xspace}
\newcommand{\mevcc}{\ensuremath{{\mathrm{\,Me\kern -0.1em V\!/}c^2}}\xspace}
\def\to                 {\ensuremath{\rightarrow}\xspace}
\newcommand{\stat}{\ensuremath{\mathrm{(stat)}}\xspace}
\newcommand{\syst}{\ensuremath{\mathrm{(syst)}}\xspace}
\def\pep2{PEP-II}

\def\CP                {\ensuremath{C\!P}\xspace}
\def\stwob{\ensuremath{\sin\! 2 \beta   }\xspace}
\def\deltat{\ensuremath{{\rm \Delta}t}\xspace}
\def\deltamd{\ensuremath{{\rm \Delta}m_d}\xspace}
\newcommand{\progtp}    [1]  {{Prog.\ Theor.\ Phys.\ {\bf #1}}}

\begin{document}  
\begin{flushleft}
\babar-PUB-05/024\\
SLAC-PUB-11321, hep-ex/0506082
\end{flushleft}

\title{
{\large  \boldmath
Measurement of Time-Dependent \CP Asymmetries and the \CP-Odd Fraction\\
in the Decay \Bztodstdst}
}
 
%
\author{B.~Aubert}
\author{R.~Barate}
\author{D.~Boutigny}
\author{F.~Couderc}
\author{Y.~Karyotakis}
\author{J.~P.~Lees}
\author{V.~Poireau}
\author{V.~Tisserand}
\author{A.~Zghiche}
\affiliation{Laboratoire de Physique des Particules, F-74941 Annecy-le-Vieux, France }
\author{E.~Grauges}
\affiliation{IFAE, Universitat Autonoma de Barcelona, E-08193 Bellaterra, Barcelona, Spain }
% ...
\collaboration{The \babar\ Collaboration}
\noaffiliation

\begin{abstract} 
We present an updated measurement of time-dependent \CP
asymmetries and the \CP-odd fraction in the 
decay $B^0 \rightarrow D^{*+}D^{*-}$ using
$232 \times 10^{6} \BB$ pairs collected
by the \babar\  detector at the
PEP-II $B$ factory.
We determine the \CP-odd fraction to be $0.125 \pm 0.044\stat \pm 0.007\syst$.
The time-dependent \CP asymmetry parameters $C_+$
and $S_+$ are determined
to be $0.06\pm 0.17\stat \pm 0.03\syst$ and $-0.75 \pm 0.25\stat \pm 0.03\syst$,  
respectively.  The
Standard Model predicts these parameters to be 0 and $-\stwob$, respectively, 
in the absence of penguin amplitude contributions. 
\end{abstract}
 
  
\maketitle

The time-dependent \CP asymmetry measurement in \Bztodstdst decay provides an
important test of the Standard Model (SM). 
In the SM,  \CP violation arises from  
a complex phase in the Cabibbo-Kobayashi-Maskawa (CKM) quark-mixing 
matrix~\cite{CKM}. Measurements of \CP asymmetries by the 
\babar~\cite{Aubert:2002rg} and BELLE~\cite{Abe:2002px} 
collaborations have firmly established this effect in the
\bpsiks decay~\cite{conjugate} and related modes that are governed by 
the $b\to\ccbar s$ transition. The \Bztodstdst decay is dominated
by  the $b\to \ccbar d$ transition.
Within the framework of the SM,
the \CP asymmetry of \Bztodstdst is related to \stwob when the 
correction due to penguin diagram contributions are neglected. The 
penguin-induced correction has been estimated in models based on
the factorization approximation and heavy quark symmetry and was
predicted to be about 2\%~\cite{Pham:1999fy}.  
A significant deviation of the
measured \stwob from the one observed in $b\to c\bar{c}s$ decays would be
evidence for a new \CP-violating interaction. The enhanced sensitivity of
\Bztodstdst to such a process arises from its much smaller SM amplitude
compared with that of the $b\to c\bar{c}s$ transition.

The \Bztodstdst decay proceeds through the \CP-even $S$ and $D$ waves and 
through the \CP-odd $P$ wave.
In this Letter, we present an improved measurement
of the \CP-odd fraction~\cite{Aubert:2003uv,Miyake:2005qb} $R_\perp$ based 
on a time-integrated one-dimensional angular analysis. We also present
an improved measurement of the time-dependent 
\CP asymmetry~\cite{Aubert:2003uv,Miyake:2005qb}, obtained 
from a combined analysis of time-dependent flavor-tagged
decays and the one-dimensional angular distribution of the decay
products. 

The data used in this analysis comprise
232 million \upsbb decays collected by the
\babar\ detector at the PEP-II storage ring. The \babar\ detector
is described in detail elsewhere~\cite{Aubert:2001tu}. 
We use a Monte Carlo (MC) simulation 
based on GEANT4~\cite{Agostinelli:2002hh}
to validate the analysis procedure and 
to study the relevant backgrounds.

We select \Bztodstdst decay by combining two charged $D^{*}$
candidates reconstructed in the modes
$\Dstarp\to\Dz\pip$ and $\Dstarp\to\Dp\piz$.
We include the $\Dstarp\Dstarm$ combinations
$(\Dz\pip, \Dzb\pim)$ and $(\Dz\pip, \Dm\piz)$,
but not $(\Dp\piz,\Dm\piz)$ because of
the smaller branching fraction and larger backgrounds.
To suppress the $\epem\to\qqbar \;(q=u,d,s,\,\rm{and}\; c)$
continuum background, we require the ratio of the
second and zeroth order Fox-Wolfram moments~\cite{Fox:1978vu} to be
less than 0.6. 

Candidates for \Dz and \Dp mesons are reconstructed in the modes
$\Dz\to\Km\pip$, $\Km\pip\piz$, $\Km\pip\pip\pim$, 
$\KS\pip\pim$ and  $\Dp\to\Km\pip\pip$, 
$\KS\pip$, $\Km\Kp\pip$. The reconstructed mass of the
\Dz(\Dp) candidate is required to be within 20\,\mevcc 
of its nominal mass~\cite{Eidelman:2004wy}, 
except for the $\Dz\to\Km\pip\piz$ candidate, where
a looser requirement of 40\,\mevcc is applied. 

The \KS candidates are reconstructed from two oppositely-charged
tracks with an invariant mass within 
20\,\mevcc of the nominal \KS mass. The $\chi^2$ probability of 
the $\pip\pim$ vertex fit must be greater than $0.1\,\%$. 
Charged kaon candidates
are required to be inconsistent with the pion hypothesis, as 
inferred from the Cherenkov angle measured by the Cherenkov 
detector and the ionization energy loss measured by the 
charged-particle tracking system~\cite{Aubert:2001tu}. 
Neutral pion candidates are formed from two photons detected in the 
electromagnetic calorimeter~\cite{Aubert:2001tu},
each with energy above 30\,\mev. The mass of the pair must be within
30\,\mevcc of the nominal \piz mass, and their summed energy 
is required to be greater than 200\,\mev. In addition,
a mass-constrained fit is applied to the \piz candidates
for further analysis.

The \Dz and \Dp candidates are subject to a mass-constrained fit
prior to the formation of the \Dstarp candidates. 
A slow \pip from \Dstarp decay is required to have a momentum in the 
\FourS center-of-mass (CM) frame less than 450\,\mevc.
A slow \piz from \Dstarp must have a momentum between 
$70$ and $450\,\mevc$ in the CM frame.
No requirement on the photon-energy sum is applied to the
\piz candidates from the \Dstarp decays.

For each \Bztodstdst candidate, we construct a 
likelihood function~\cite{Aubert:2002vn}
$\mathcal{L}_{\rm{mass}}$ from the masses and mass uncertainties
of the $D$ and \Dstar candidates. The 
likelihood $\mathcal{L}_{\rm{mass}}$ 
is calculated as the product of the likelihoods for the
$D$ and \Dstar candidates.
The $D$ mass resolution is modeled by a Gaussian whose variance
is determined on a candidate-by-candidate basis.
The \Dstar-$D$ mass difference resolution is modeled by a double-Gaussian
distribution whose parameters are determined from simulated events.
The values of $\mathcal{L}_{\rm{mass}}$
and the difference of the \Bz candidate energy $E_B$ from 
the beam energy $E_{\rm{Beam}}$, $\Delta E\equiv E_B-E_{\rm{Beam}}$, 
in the \FourS CM frame are used to reduce the
combinatoric background further. From the simulated events, 
the maximum allowed values of 
$-\ln\mathcal{L}_{\rm{mass}}$ and $|\Delta E|$ are optimized  
for each individual final state to obtain the 
highest expected signal significance using the previously
measured \Bztodstdst branching fraction~\cite{Aubert:2003uv}.

The energy-substituted mass, $m_{\rm{ES}}\equiv \sqrt{E^2_{\rm{Beam}}-p^{*2}_B}$, 
where $p^*_B$ is
the $B^0$ candidate momentum in the \FourS CM frame, is used
to extract the signal yield from the events satisfying the 
aforementioned selection. We select the \Bz
candidates that have $m_{\rm{ES}}\ge5.23\,\gevcc$. In cases where
more than one \Bz candidate is reconstructed in an event, the
candidate with the smallest  value of $-\ln\mathcal{L}_{\rm{mass}}$ is chosen.
A fit to the $m_{\rm{ES}}$ distribution 
with a probability density function (PDF) given by the sum of a Gaussian
shape for the signal and an ARGUS~\cite{Albrecht:1990cs} function for 
the background yields
$391\pm28\stat$  signal events. In the region of $m_{\rm{ES}}> 5.27\,\gevcc$,
the signal purity is approximately 70\,\%.

In the transversity basis~\cite{Dunietz:1990cj}, 
we define the following three angles:
the angle $\theta_1$ between the momentum of the
slow pion from the \Dstarm 
and the opposite direction of flight of the \Dstarp in the \Dstarm rest frame;
the polar angle $\theta_{\rm tr}$ and  azimuthal angle $\phi_{\rm tr}$
of the slow pion from the \Dstarp defined in the \Dstarp rest frame,
where the opposite direction of flight of the  \Dstarm
is chosen as the $x$-axis, and the $z$-axis is defined
as the normal to the \Dstarm decay plane.

The time-dependent
angular distribution of the decay products is given in Ref.~\cite{penguin}.
Taking into account the detector angular acceptance efficiency
and integrating over the decay time and the angles $\theta_1$ and
$\phi_{\rm tr}$, we obtain a one-dimensional differential decay rate:
\begin{eqnarray}
\frac{1}{\Gamma}  \frac{d \Gamma}{d \cos \theta_{\rm tr}} & = &
  \frac{9}{32\pi} \left[ (1-R_\perp) \sin^2 \theta_{\rm tr}  \right. \nonumber \\
& \times & \left\{ \frac{1+\alpha}{2} I_{0}(\cos\theta_{\rm tr}) +
\frac{1-\alpha}{2} I_{\parallel}(\cos\theta_{\rm tr})
\right\} \nonumber \\
& + & \left. 2 R_\perp \cos^2 \theta_{\rm tr}
\times I_{\perp}(\cos\theta_{\rm tr}) \right],
\label{AngDisArt}
 \end{eqnarray}
where
$R_\perp = |A_\perp|^2 / (|A_0|^2 + |A_\parallel|^2 + |A_\perp|^2)$,
$\alpha = (|A_{0}|^2  - |A_{\parallel}|^2) / (|A_0|^2 + |A_{\parallel}|^2)$,
$A_0$ is the amplitude for longitudinally polarized $D^*$ mesons, $A_{\parallel}$
and $A_{\perp}$ are the amplitudes for parallel and perpendicular transversely 
polarized $D^*$ mesons.
The three efficiency moments, $I_k$ $(k=0, \parallel, \perp)$, are defined as
\begin{eqnarray}
\displaystyle
I_{k}(\cos\theta_{\rm tr}) = \int \! d \! \cos \theta_1 \, d \phi_{\rm tr}
\; g_k(\theta_1, \phi_{\rm tr}) \,\varepsilon(\theta_1,\theta_{\rm tr},\phi_{\rm tr}),
\label{moments}
\end{eqnarray}
where $g_0 = 4\cos^2\theta_1 \cos^2\phi_{\rm tr}$, $g_{||} = 2\sin^2\theta_1 \sin^2\phi_{\rm tr}$,
$g_{\perp} = \sin^2\theta_1 $, and $\varepsilon$ is the detector efficiency.
The efficiency moments are parameterized as second-order even polynomials 
of $\cos\theta_{\rm tr}$. Their parameter values are determined from the
MC and are subsequently fixed in the likelihood fit to the differential
decay distribution of $\cos\theta_{\rm tr}$. In fact, the three $I_{k}$
functions deviate only slightly from a constant, making the 
distribution, Eq.~\ref{AngDisArt}, nearly independent of the amplitude 
ratio $\alpha$. 

\begin{figure}[!ht]
\begin{center}
\scalebox{0.45}{\includegraphics{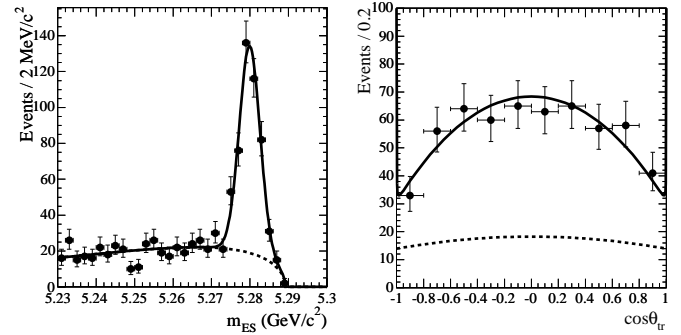}}
\caption{ Measured distribution of $m_{\rm ES}$ (left)
and of $\cos\theta_{\rm tr}$
in the region $\mes > 5.27\,\gevcc$ (right).
The solid line is the projection of the fit result.  
The dotted line represents the background component.
}
\label{fig:datafit_rt}
\end{center}
\end{figure}

The \CP-odd fraction $R_\perp$ is measured in a simultaneous
unbinned maximum likelihood fit 
to the $\cos\theta_{\rm tr}$ and the $m_{\rm ES}$ distribution.
The background shape is modeled
as an even second-order polynomial in $\cos\theta_{\rm tr}$, while
the signal PDF is given by Eq.~\ref{AngDisArt}. The finite 
detector resolution of the $\theta_{\rm tr}$ measurement 
is modeled as a double Gaussian plus a small tail component
that accounts for misreconstructed events. The parameterization
of the $\theta_{\rm tr}$ resolution function is fixed from the MC simulation 
and subsequently 
used to convolve the signal PDF in the maximum likelihood fit. Since the 
angle $\theta_{\rm tr}$ is calculated with the slow pion from the 
\Dstarp, we categorize events into three types: 
$\Dstarp\Dstarm \to (\Dz\pip,\Dzb\pim)$,
$(\Dz\pip,\Dm\piz)$, and $(\Dp\piz,\Dzb\pim)$, each with different
signal-fraction parameters in the likelihood fit. Their angular efficiency moments 
and $\cos\theta_{\rm tr}$ resolutions are also separately determined from the
MC simulation. The other parameters determined in the likelihood fit are the
$\cos\theta_{\rm tr}$ background-shape parameter, three $m_{\rm ES}$ parameters
($\sigma$ and mean of the signal Gaussian, and the ARGUS shape parameter $\kappa$),
as well as $R_\perp$. The fit to the data yields
\begin{equation}
\displaystyle
R_\perp = 0.125 \pm 0.044\stat \pm 0.007\syst.
\end{equation}
The projections of the fitted result onto $m_{\rm ES}$ and
$\cos\theta_{\rm tr}$ are shown in 
Fig.~\ref{fig:datafit_rt}.

In the fit described above, the value of $\alpha$ is fixed to zero. 
We estimate the
corresponding systematic uncertainty by varying its value from 
$-1$ to $+1$ and find negligible change (less than 0.002) in the fitted 
value of $R_\perp$.
Other systematic uncertainties arise from the parameterization of the 
angular resolution, the determination of the efficiency moments, and
the background parameterization. The total systematic uncertainty on
$R_\perp$ is 0.007, significantly smaller than the statistical error.

We subsequently perform a combined analysis of the $\cos\theta_{\rm tr}$
distribution and the time dependence to extract the time-dependent \CP asymmetry,
using the event sample described previously. We use information
from the other $B$ meson in the event to tag the initial flavor
of the fully reconstructed \Bztodstdst candidate. 

The decay rate $f_+ (f_-)$ for a neutral $B$ meson accompanied by  a $B^{0}
(\Bzb)$ tag is given by
\begin{eqnarray}
f_\pm(\deltat,\cos\theta_{\rm tr}) \propto
{\rm e}^{ - | \deltat |/\tau_{B^0} }
 \Bigl\{ G(1\mp\Delta\omega) \mp (1-2\omega)\nonumber \\
 \left[
 F\sin{ (\deltamd  \deltat) }
+ H\cos{ (\deltamd  \deltat) }  \right]  \Bigr\},
 \label{eq:sincos}
\end{eqnarray}
where $\Delta t = t_{\rm rec} - t_{\rm tag}$ is the difference between
the proper decay time of the reconstructed $B$ meson ($B_{\rm rec}$) and
that of the tagging $B$ meson ($B_{\rm tag}$),
$\tau_{\Bz}$ is the \Bz lifetime, and \deltamd is the mass difference 
determined from the \Bz-\Bzb oscillation frequency~\cite{Eidelman:2004wy}.
The average mistag probability $\omega$ describes
the effect of incorrect tags, and $\Delta\omega$
is the difference between 
the mistag rate for $\Bz$ and $\Bzb$. The $G$, $F$ and $H$ coefficients
are defined as:
\begin{eqnarray}
G &=&  (1-R_\perp) \sin^2\theta_{\rm tr}
   +    2  R_\perp \cos^2\theta_{\rm tr},\nonumber \\
F &=&  (1-R_\perp)S_+ \sin^2\theta_{\rm tr}-
2R_\perp S_\perp\cos^2\theta_{\rm tr},\\
H &=&  (1-R_\perp)C_+ \sin^2\theta_{\rm tr}+
2R_\perp C_\perp\cos^2\theta_{\rm tr},\nonumber 
\end{eqnarray}
where we allow the three transversity amplitudes to have different
$\lambda_k=(q/p)(\bar{A}_k/A_k)$ $(k=0, \parallel, \perp)$~\cite{penguin}
due to 
possibly different penguin-to-tree amplitude ratios, and define the
\CP asymmetry $C_k=1-|\lambda_k|^2/1+|\lambda_k|^2$,
$S_k=2\Im(\lambda_k)/1+|\lambda_k|^2$.
Here we also have:
\begin{equation}
C_+=\frac{C_\parallel |A_\parallel|^2+C_0|A_0|^2}
{|A_\parallel|^2+|A_0|^2},
S_+=\frac{S_\parallel |A_\parallel|^2+S_0|A_0|^2}
{|A_\parallel|^2+|A_0|^2}.
\end{equation}
In the absence of penguin contributions, we expect that
$C_0=C_\parallel=C_\perp=0$, and $S_0=S_\parallel=S_\perp=-\stwob$.

In Eq.~\ref{eq:sincos}, the small angular acceptance effects are not 
incorporated, but absorbed into the "effective" value of
$R_\perp$, which is left free to vary in the final fit. No
bias is seen in the resulting values of $C_+$, $C_\perp$, $S_+$, 
and $S_\perp$ in MC simulation. 

The technique used to measure the \CP asymmetry is analogous to previous
\babar\ measurements as described in Ref.~\cite{Aubert:2004zt}. 
Only events with 
a $\Delta t$ uncertainty less than $2.5\,\mbox{ps}$ and a measured 
$|\Delta t|$ less than $20\,\mbox{ps}$ are accepted. 
We performed a simultaneous unbinned maximum likelihood fit to
the $\cos\theta_{\rm tr}$, $\Delta t$, and $m_{\rm ES}$ 
distributions 
to extract the \CP asymmetry. The signal PDF in $\theta_{\rm tr}$ and
$\Delta t$ is given by Eq.~\ref{eq:sincos}.
The signal mistag probability is determined 
from a sample of neutral $B$ decays to flavor eigenstates, $B_{\rm flav}$.
In the likelihood fit, the expression in Eq.~\ref{eq:sincos} is  
convolved with an empirical $\Delta t$ resolution function determined
from the $B_{\rm flav}$ sample. The $\theta_{\rm tr}$ resolution
is accounted for in the same way as described previously. 

The background $\Delta t$ distributions are parameterized with an empirical 
description that includes prompt and non-prompt components. 
We allow the non-prompt
background to have two free parameters, $C_{\rm eff}$ and 
$S_{\rm eff}$, the effective \CP asymmetries, in the likelihood fit.
The background shape
in $\theta_{\rm tr}$ is modeled as an even second-order polynomial in 
$\cos\theta_{\rm tr}$, much as it is in the time-integrated angular analysis.

\begin{figure}[!ht]
\begin{center}
\scalebox{0.4}{\includegraphics{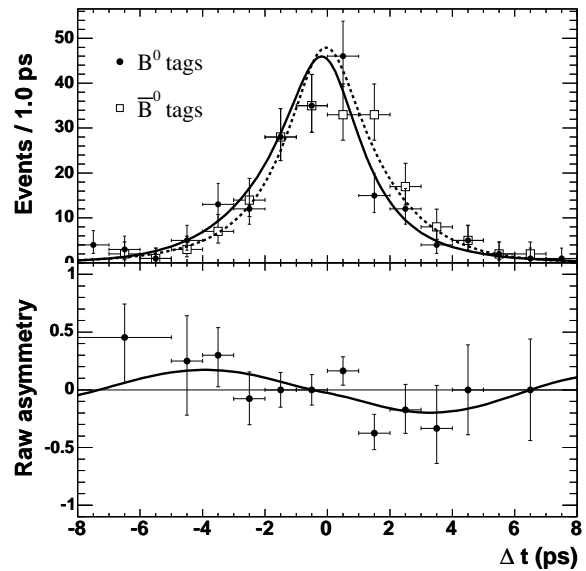}}
\caption{ 
From top to bottom: the distribution of $\Delta t$ in the region  
$\mes > 5.27\,\gevcc$ for \Bz (\Bzb) tag candidates,
and the raw asymmetry
$(N_{\Bz}-N_{\Bzb})/(N_{\Bz}+N_{\Bzb})$, as functions of \deltat . 
In the upper plot the
solid (dashed) curves represent the fit projections in $\Delta t$ for 
\Bz (\Bzb) tags.
}
\label{fig:datafit_cp}
\end{center}
\end{figure}

The fit to the data yields
\begin{eqnarray}
C_+&=&0.06\pm 0.17\stat\pm 0.03\syst,\nonumber\\
C_\perp&=&-0.20\pm 0.96\stat\pm 0.11\syst,\nonumber\\
S_+&=&-0.75\pm 0.25\stat\pm 0.03\syst,\nonumber\\
S_\perp&=&-1.75\pm 1.78\stat\pm 0.22\syst.
\end{eqnarray}
Fig.~\ref{fig:datafit_cp} shows the $\Delta t$ distributions
and asymmetries in yields between $B^0$ and $\Bzb$ 
tags, overlaid with the projection of the likelihood fit result.
Because the \CP-odd fraction is small, we have rather large
statistical uncertainties for the measured $C_\perp$ and $S_\perp$
values. For comparison, we 
repeat the fit with the assumption that both \CP-even
and \CP-odd states have the same \CP asymmetry. We find
that $C_+=C_\perp=0.03\pm 0.13\stat\pm 0.02\syst$, 
and $S_+=S_\perp=-0.69\pm 0.23\stat\pm 0.03\syst$. In both cases,
the effective \CP asymmetries in the background are found to be
consistent with zero within the statistical uncertainties. 

The systematic uncertainties on $C_+$, $C_\perp$, $S_+$ and $S_\perp$ 
arise 
from the amount of possible backgrounds that tend to peak
under the signal and their
\CP asymmetry, the assumed parameterization of the 
$\Delta t$ resolution function, the possible differences between the 
$B_{\rm flav}$ and $B_{\CP}$ mistag fractions, knowledge
of the event-by-event beam-spot position, and the possible
interference between  the suppressed 
$\bar{b}\to\bar{u}c\bar{d}$ amplitude and the favored
$b\to c\bar{u}d$ amplitude for some tag-side decays~\cite{Long:2003wq}.
It also includes the systematic uncertainties from the finite MC 
sample used to verify the fitting method. In general, all of 
the systematic
uncertainties are found to be much smaller than the statistical
uncertainties.

In summary, we have reported measurements of the \CP-odd fraction
and time-dependent \CP
asymmetries for the decay
\Bztodstdst.  The measurement
supersedes the previous \babar\ result~\cite{Aubert:2003uv},
with more than $50\,\%$ reduction in the statistical uncertainty, and
indicates that \Bztodstdst is mostly \CP-even.
The time-dependent asymmetries
are found to be consistent with the SM predictions
within the statistical uncertainty.

We are grateful for the excellent luminosity and machine conditions
provided by our \pep2\ colleagues, 
and for the substantial dedicated effort from
the computing organizations that support \babar.
The collaborating institutions wish to thank 
SLAC for its support and kind hospitality. 
This work is supported by
DOE
and NSF (USA),
NSERC (Canada),
IHEP (China),
CEA and
CNRS-IN2P3
(France),
BMBF and DFG
(Germany),
INFN (Italy),
FOM (The Netherlands),
NFR (Norway),
MIST (Russia), and
PPARC (United Kingdom). 
Individuals have received support from CONACyT (Mexico), A.~P.~Sloan Foundation, 
Research Corporation,
and Alexander von Humboldt Foundation.


\begin{thebibliography}{99}
\bibitem{conjugate}
We imply charged conjugate modes throughout the paper.
\end{thebibliography}
\end{document}